\documentclass[10pt, journal]{IEEEtran}

\usepackage{graphicx, enumerate}
\usepackage{cite,array, bm}
\usepackage{dsfont, epstopdf}
\usepackage{mathrsfs, subfigure, color}
\usepackage{amssymb, multirow}
\usepackage[cmex10]{amsmath}
\usepackage{subeqnarray, tabularx}
\usepackage{cases, stmaryrd}
\usepackage{algpseudocode}
\usepackage[ruled,vlined,linesnumbered]{algorithm2e}
\usepackage[utf8]{inputenc}
\usepackage{threeparttable}
\usepackage{stackengine}
\usepackage{verbatim}

\DeclareFontFamily{U}{FdSymbolC}{}
\DeclareFontShape{U}{FdSymbolC}{m}{n}{<-> s * FdSymbolC-Book}{}
\DeclareSymbolFont{fdarrows}{U}{FdSymbolC}{m}{n}
\DeclareMathSymbol{\vDdash}{\mathrel}{fdarrows}{254}

\DeclareFontFamily{U}{FdSymbolD}{}
\DeclareFontShape{U}{FdSymbolD}{m}{n}{<-> s * FdSymbolD-Book}{}
\DeclareSymbolFont{fdnarrows}{U}{FdSymbolD}{m}{n}
\DeclareMathSymbol{\nvDdash}{\mathrel}{fdnarrows}{254}
\makeatother

\newtheorem{assumption}{Assumption}
\newtheorem{theorem}{Theorem}

\newtheorem{lemma}{Lemma}
\newtheorem{proposition}{Proposition}

\definecolor{bluencs}{rgb}{0.0, 0.53, 0.74}

\begin{document}

\title{\LARGE \bf Optimal Resource Scheduling and Allocation \\ in Distributed Computing Systems
\thanks{This work was supported by the H2020 ERC Consolidator Grant L2C (Grant 864017), the CHIST-ERA 2018 project DRUID-NET, the Walloon Region and the Innoviris Foundation.}
}

\author{Wei Ren, Eleftherios Vlahakis, Nikolaos Athanasopoulos and Rapha\"el Jungers
\thanks{W. Ren and R. Jungers are with ICTEAM institute, UCLouvain, 1348 Louvain-la-Neuve, Belgium. E. Vlahakis and N. Athanasopoulos are with School of Electronics, Electrical Engineering and Computer Science, Queen’s University Belfast, Northern Ireland, UK. Email: \texttt{\small w.ren@uclouvain.be, e.vlahakis@qub.ac.uk, n.athanasopoulos@qub.ac.uk, \\ raphael.jungers@uclouvain.be}.
}}

\maketitle

\begin{abstract}
The essence of distributed computing systems is how to schedule incoming requests and how to allocate all computing nodes to minimize both time and computation costs. In this paper, we propose a cost-aware optimal scheduling and allocation strategy for distributed computing systems while minimizing the cost function including response time and service cost. First, based on the proposed cost function, we derive the optimal request scheduling policy and the optimal resource allocation policy synchronously. Second, considering the effects of incoming requests on the scheduling policy, the additive increase multiplicative decrease (AIMD) mechanism is implemented to model the relation between the request arrival and scheduling. In particular, the AIMD parameters can be designed such that the derived optimal strategy is still valid. Finally, a numerical example is presented to illustrate the derived results.
\end{abstract}

\section{Introduction}
\label{sec-intro}

With the advances in networking and hardware technology, distributed computing is increasingly demanded in many applications like multi-agent systems \cite{Woolridge2001introduction}, Internet multimedia traffic \cite{Tang2014dynamic} and network systems \cite{Kshemkalyani2011distributed}. For this purpose, distributed computing systems have been proposed to tie together the power of large number of resources distributed across networks \cite{Hussain2013survey}. In particular, scalability, reliability, information sharing and information exchange from remote sources are main motivations for the users of distributed systems \cite{Amir2000opportunity}. A distributed computing system is composed of a set of computing/processing nodes among which communication links are established. From different perspectives of computing systems, many research topics have been investigated \cite{Huang2020supervisory, Kimura2020packet, Pentelas2020network, Herrera2016resource}, such as power management, request scheduling, resource allocation and system reliability, thereby resulting in different models, algorithms and software tools.

Among these research topics, resource allocation is widely recognized to be an important research problem \cite{Hussain2013survey, Huang2020supervisory}. The resource management mechanism determines the efficiency of the used resources and guarantees the Quality of Service (QoS) provided to the users. In this respect, resource allocation mechanisms include scheduling and allocation techniques to assign resources (sub-)optimally while considering the task characteristics and QoS requirements. The scheduling technique determines how to distribute (or even allocate) requests among various nodes, whereas the resource allocation pertains to the control of available resources that are provisioned to requests entering a computing system. In order to deal with the resource allocation problems in different environments like cluster, cloud, grid computing systems \cite{Hussain2013survey}, many resource allocation models and mechanisms have been proposed, such as proportional-share scheduling, market-based and auction-based mechanisms \cite{Wu2006dynamic, Sandholm2010dynamic}. Among these mechanisms, complete tandem queuing models address the request arrival process and enable to analyze the QoS performance \cite{Le2008tandem}. In particular, for the two-station model, the first station is a buffer to store the incoming requests before being allocated to all computing nodes in the second station. Therefore, how to allocate the requests from the first station into the second station and how to assign all computing nodes to deal with the allocated requests are two essential problems interacting with each other. However, many existing works consider these two problems separately \cite{Presti1996bounds, Le2008tandem} or do not involve the QoS performance analysis \cite{Van2017optimal}. For instance, different QoS performances, including loss rates and average delays, were considered in \cite{Le2008tandem}, whereas the system properties, such as station lengths and structure properties, were studied in \cite{Presti1996bounds, Van2017optimal}. Although, in \cite{Vlahakis2021AIMD}, a simultaneous scheduling and resource allocation solution with AIMD dynamics is designed, the proposed scheme is not tuned via an optimality criterion. In many existing works, the arrival and processing of the requests are generally modeled into stochastic processes like Poisson process, which is adapted in this paper.

Motivated by the above discussion, in this paper we investigate how to synchronously and dynamically select request scheduling and resource allocation of distributed computing systems so that a cost function associated with QoS metrics is minimised. That is, we aim to balance the request scheduling rate defined as the number of requests allocated by a dispatcher per unit time and the service capacity defined as the number of requests processed per unit time. To be specific, based on the response time that each request experiences and the power consumption of each computing node \cite{Li2020performance}, we first propose the cost function to evaluate the QoS performance. Second, the optimization problem is constructed by combining the cost function and the constraints from the system setup. Finally, using optimization theory, the optimal request scheduling and resource allocation strategy is derived synchronously for all computing nodes. Since the price of each computing node is introduced to determine which computing nodes to be allocated to the incoming requests, the derived optimal strategy is dynamic.

Since the request scheduling is not arbitrary and needs to balance the request arrival and processing, we next investigate the mechanism of the request scheduling and how to guarantee the validity of the derived optimal strategy in this case. In particular, we apply the well-known AIMD-based mechanism \cite{Studli2017aimd, Ravasio2021distributed, Fan2020optimized} to model the effects of the request arrival on the scheduling rate. The AIMD-based mechanism was initially proposed to prevent congestion in computer networks and subsequently applied in many fields \cite{Ucer2019analysis, Fan2020optimized} including resource allocation. The advantage of the AIMD-based mechanism lies in providing a decentralized strategy for the scenario where the communication is limited and the privacy preservation is required \cite{Corless2016aimd}. However, there are few works on the AIMD-based mechanisms to deal with scheduling problems. In this paper, we establish the convergence of the scheduling rates under the AIMD-based mechanism. To validate the derived optimal strategy under the AIMD-based mechanism, the AIMD parameters are designed such that the convergence of the scheduling rates is to the derived optimal scheduling rates. To the best of our knowledge, the AIMD parameters are designed and optimized explicitly for the first time in this paper, whereas only the properties of AIMD parameters/matrix are applied in existing works \cite{Corless2016aimd, Ucer2019analysis, Fan2020optimized} where AIMD parameters are given \emph{a priori} or can be chosen randomly. In conclusion, we combine the optimization problem with the AIMD-based mechanism to derive an optimal scheduling and allocation strategy for distributed computing systems.

The remainder of this paper is structured as follows. Section \ref{sec-notepre} introduces the system setup and the problem to be studied. The cost-aware optimal strategy for the request scheduling and resource allocation is proposed in Section \ref{sec-optimalschedul}. The AIMD-based scheduling policy is developed in Section \ref{sec-AIMDschedule}. A numerical example is given in Section \ref{sec-example}. Conclusion and future work are presented in Section \ref{sec-conclusion}.

\emph{Notation.} $\mathbb{R}:=(-\infty, +\infty)$; $\mathbb{R}^{+}:=[0, +\infty)$; $\mathbb{N}:=\{0, 1, \ldots\}$; $\mathbb{N}^{+}:=\{1, 2, \ldots\}$. $\mathbb{R}^{n}$ denotes the $n$-dimensional Euclidean space. Given a function $f: \mathbb{R}^{n}\rightarrow\mathbb{R}$ and $x:=(x_{1}, \ldots, x_{n})\in\mathbb{R}^{n}$, we denote by $\nabla f(x)$ the derivative vector $(\partial f(x)/\partial x_{1}, \ldots, \partial f(x)/\partial x_{n})$.

\section{System Setup and Problem Formulation}
\label{sec-notepre}

In this section, we describe our model and system setup, define some notations, and state our assumptions.

\subsection{System Description}
\label{subsec-notation}

\begin{figure}[!t]
\begin{center}
\begin{picture}(55, 110)
\put(-60, -12){\resizebox{55mm}{40mm}{\includegraphics[width=2.5in]{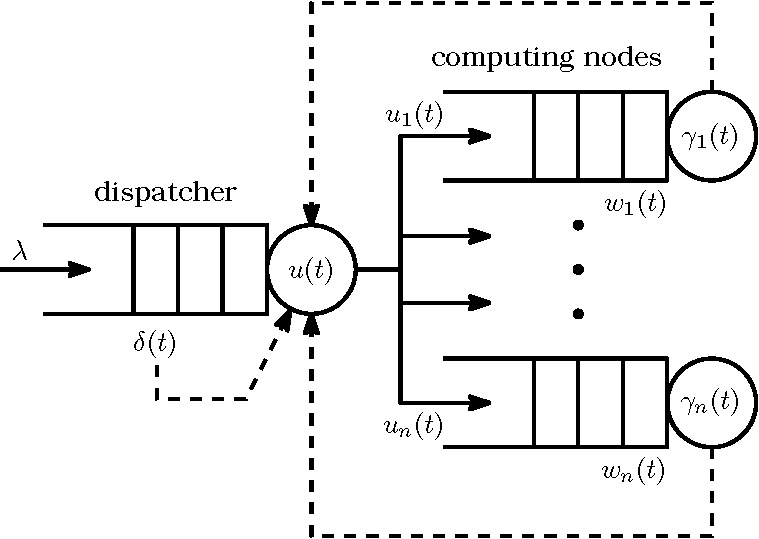}}}
\end{picture}
\end{center}
\caption{Illustration of the model for request scheduling and resource allocation.}
\label{fig-1}
\end{figure}

We consider the problem of scheduling an arrival request process at a dispatcher or load balancer in distributed computing systems such as cloud-centric networks, which tends to experience Poisson bursts of user traffic \cite{Ranjan2008high}. The request scheduling and resource allocation of distributed computing systems are illustrated in Fig. \ref{fig-1}. A request is an individual demand for computing resources provided by a computing node. The request arrival is modeled as a Poisson process with an arrival rate $\lambda\in\mathbb{R}^{+}$; see also \cite{Van2017optimal, Presti1996bounds, Le2008tandem}. Also, we assume that the arrival request process consists of a stream of transactions, each of which is to be allocated to a single computing node defined as a physical or virtual machine and environment. This reflects practical architectures like MapReduce \cite{Dean2008mapreduce}, in which multiple computing nodes are allocated via the map function at the dispatcher.

After receiving the incoming requests, the dispatcher schedules these requests into multiple computing nodes. The desired case is that the dispatcher schedules all received requests; otherwise, some requests are stored in the dispatcher and queued for scheduling. To describe the number of all queued requests, we introduce the variable $\delta(t)\in\mathbb{R}^{+}$ to denote the backlog of the dispatcher. Let the number of all computing nodes be $n\in\mathbb{N}^{+}$. Each request is scheduled to a computing node, which scales its service capacity to minimize the overall processing time and cost. For the $i$-th computing node with $i\in\mathcal{N}:=\{1, \ldots, n\}$, $u_{i}(t)\in\mathbb{R}^{+}$ is its scheduling rate from the dispatcher, and $\gamma_{i}(t)\in\mathbb{R}^{+}$ is its service rate, i.e., the number of arrival requests that can be serviced by the $i$-th computing node per unit time. Both $u_{i}(t)$ and $\gamma_{i}(t)$ are design parameters. It is clear that $u(t):=\sum^{n}_{i=1}u_{i}(t)\leq\lambda$. If the scheduling rate $u_{i}(t)$ is larger than the service rate $\gamma_{i}(t)$, the $i$-th computing node may store some requests that cannot be served in time. In this respect, similar to the dispatcher, the variable $w_{i}(t)\in\mathbb{R}^{+}$ is introduced to denote the backlog of the $i$-th computing node. Note that not all computing nodes need to be implemented, which implies that both $u_{i}(t)$ and $\gamma_{i}(t)$ can be zero.

For the request scheduling and resource allocation as in Fig. \ref{fig-1}, the dynamics of $\delta(t)$ and $w_{i}(t)$ is derived below \cite{Presti1996bounds, Vlahakis2021AIMD}:
\begin{align}
\label{eqn-1}
\begin{aligned}
\dot{\delta}(t)&=\lambda-u(t), \\
\dot{w}_{i}(t)&=u_{i}(t)-\gamma_{i}(t).
\end{aligned}
\end{align}
If all incoming requests can be scheduled well, then $\delta(t)$ can be zero at some time instant. In addition, $\delta(t)$ can be lower/upper bounded in practice. These cases may affect the scheduling rates such that the all scheduling rates experience a jump, which can be modeled as an event-triggered system. A well-known model for this case is AIMD dynamics \cite{Corless2016aimd}, and will be introduced in Section \ref{sec-AIMDschedule}. On the other hand, since the scheduling and service rates can be designed, we can require $u_{i}(t)$ to be smaller than $\gamma_{i}(t)$ such that $w_{i}(t)$ does not need to be considered. This requirement is also reasonable in terms of the service cost \cite{Tang2014dynamic}. To be specific, the storage of all requests in each computing node results in memory cost and an increase in the service response time, thereby resulting in the inefficiency of the resource allocation mechanism. From the above discussion, the following assumption is made.

\begin{assumption}
\label{asp-1}
For the request scheduling and allocation system, the following holds.
\begin{enumerate}
  \item For each $i\in\mathcal{N}$, $\gamma_{i}(t)$ is upper bounded by a constant $\gamma^{\max}_{i}\in\mathbb{R}^{+}$.
  \item For each $i\in\mathcal{N}$ and all $t\in\mathbb{R}^{+}$, $\gamma_{i}(t)>u_{i}(t)$.
\end{enumerate}
\end{assumption}

In Assumption \ref{asp-1}, the first item comes from the nature of each computing node, whose computation capacity is definitely limited. The second item can be imposed via the scheduling and allocation strategy to be designed.

\subsection{Cost Function for the Scheduling and Allocation}
\label{subsec-approbisimu}

In distributed computing systems, the request scheduling problem involves how to redirect incoming requests to all computing nodes, whereas the resource allocation problem pertains to how to rearrange the resource of each computing node such that the scheduled requests can be served quickly with a minimal cost. Therefore, our goal is to propose a cost-aware optimal request scheduling and resource allocation strategy for distributed computing systems.

To evaluate the scheduling and allocation, we first consider the QoS performance. Here, we consider the following average mean response time \cite{Tang2014dynamic} as the QoS performance:
\begin{align}
\label{eqn-2}
\mathcal{T}:=\sum_{i=1}^{n} \frac{u_{i}}{\lambda}\frac{1}{\gamma_{i}-u_{i}},
\end{align}
which is the average mean of all response times $(\gamma_{i}-u_{i})^{-1}$ with $i\in\mathcal{N}$. From \eqref{eqn-2}, we can see the reasonability of item 2) in Assumption \ref{alg-1}. Since $\gamma_{i}>u_{i}\geq0$ holds from Assumption \ref{asp-1}, we can check that $\mathcal{T}$ is convex with respect to each $\gamma_{i}$ satisfying $\gamma_{i}>u_{i}$, and convex with respect to each $u_{i}$ satisfying $0<u_{i}<\gamma_{i}$. Here, we point out that besides response times, both reliability and security can be treated as the QoS performances \cite{Li2020performance, Wu2006dynamic}.

In addition to the QoS performance, we need to consider the cost of all computing nodes. Here, we investigate the power consumption via the total service cost defined below:
\begin{align}
\label{eqn-3}
\mathcal{C}:=\sum_{i=1}^{n}\lambda^{-1}u_{i}\varphi_{i}(\gamma_{i}),
\end{align}
where $\varphi_{i}(\gamma_{i}): \mathbb{R}^{+}\rightarrow\mathbb{R}^{+}$ is the service cost for each computing node. The service cost $\varphi_{i}(\gamma_{i})$ includes the routing cost and the computing cost. The routing cost is assumed to be a constant, whereas the computing cost is related to and non-decreasing in the service rate $\gamma_{i}$. That is, the computing cost will increase if more computing resources are requested by each computing node. The computing cost includes the cost of the power used by each computing node and the memory costs required by each computing node. In this respect, for each computing node, its service cost is defined as follows:
\begin{align}
\label{eqn-4}
\varphi_{i}(\gamma_{i}):=a_{i}\gamma^{b_{i}}_{i}+c_{i}\gamma_{i}+d_{i}, \quad \forall i\in\mathcal{N},
\end{align}
where $a_{i}, c_{i}, d_{i}>0$ and $b_{i}>1$. It is easy to check that $\varphi_{i}(\gamma_{i})$ is convex. In addition, the derivative of $\varphi_{i}(\gamma_{i})$ with respect to $\gamma_{i}$, that is, $\varphi^{\prime}_{i}(\gamma_{i}):=a_{i}b_{i}\gamma^{b_{i}-1}_{i}+c_{i}$, is positive and increasing with respect to $\gamma_{i}\in\mathbb{R}^{+}$. In \eqref{eqn-5}, the first item is the power cost which is monomial in the service rate \cite{Chen2015interaction}; the second item is the cost of processor and storage memory; the last item is the routing cost. Since the service rate is upper bounded in practical systems, we can define $\varphi_{i}(\gamma_{i})$ as $\infty$ if $\gamma_{i}>\gamma^{\max}_{i}$; see, e.g., \cite{Tang2014dynamic} and some physical applications like Amazon elastic compute cloud (EC2) \cite{Yi2010reducing}.

Combining the QoS performance \eqref{eqn-2} and the power consumption \eqref{eqn-3}, we yield the following cost function:
\begin{align}
\label{eqn-5}
\mathcal{J}:=\sum_{i=1}^{n} \frac{u_{i}}{\lambda}\frac{1}{\gamma_{i}-u_{i}}+K\sum_{i=1}^{n}\frac{u_{i}}{\lambda}\varphi_{i}(\gamma_{i}),
\end{align}
where $K>0$ is a fixed weight and $\varphi_{i}$ is defined in \eqref{eqn-4}.

\subsection{Problem Formulation}

Under Assumption \ref{asp-1}, our objective is to find the scheduling rate vector $u:=(u_{1}, u_{2}, \ldots, u_{n})\in\mathbb{R}^{n}$ and the service rate vector $\gamma:=(\gamma_{1}, \gamma_{2}, \ldots, \gamma_{n})\in\mathbb{R}^{n}$ of all computing nodes such that the cost function $\mathcal{J}$ in \eqref{eqn-5} is minimized. To this end, the following optimization problem is formulated.
\begin{subequations}
\label{eqn-6}
\begin{align}
\label{eqn-6a}
\min&\quad \sum_{i=1}^{n}\lambda^{-1}u_{i}\left((\gamma_{i}-u_{i})^{-1}+K\varphi_{i}(\gamma_{i})\right) \\
\label{eqn-6b}
\text{s.t.}& \quad \sum_{i=1}^{n}u_{i}=\lambda,  \\
\label{eqn-6c}
&\quad \gamma_{i}\leq\gamma^{\max}_{i},\quad \forall i\in\mathcal{N}, \\
\label{eqn-6d}
&\quad \gamma_{i}>u_{i}\geq0,\quad \forall i\in\mathcal{N}.
\end{align}
\end{subequations}
In this optimization problem, we aim to minimize the cost function \eqref{eqn-6a}, which equals to \eqref{eqn-5}. The constraint \eqref{eqn-6b} naturally comes from the system setup, and the ideal case is $\sum_{i=1}^{n}u_{i}=\lambda$. The constraints \eqref{eqn-6c}-\eqref{eqn-6d} come from Assumption \ref{asp-1}.

From the optimization problem \eqref{eqn-6}, the backlog of the dispatcher and all computing nodes is not involved. Therefore, to deal with the optimization problem \eqref{eqn-6}, we first consider the backlog-free case and the optimal strategy of which is derived in Section \ref{sec-optimalschedul}. Due to the constraint \eqref{eqn-6d} and the discussion in Section \ref{subsec-notation}, the backlog of all computing nodes does not need to be considered. To address the backlog of the dispatcher, we further investigate the mechanism of the request scheduling in Section \ref{sec-AIMDschedule}. In particular, the AIMD mechanism is implemented and the AIMD parameters are designed explicitly to guarantee that the derived optimal strategy is still valid.

\section{Optimal Scheduling and Allocation}
\label{sec-optimalschedul}

In this section, we establish the optimal scheduling and service rates for \eqref{eqn-6}. Since not all computing nodes need to be activated, we first focus on how to choose the computing nodes to be activated. Based on the cost function \eqref{eqn-6a}, we first introduce the following variable for each computing node:
\begin{align}
\label{eqn-7}
\theta_{i}:=\min_{0<\gamma\leq\gamma^{\max}_{i}}\{\gamma^{-1}+K\varphi_{i}(\gamma)\}, \quad \forall  i\in\mathcal{N},
\end{align}
which can be treated as the price of each computing node \cite{Tang2014dynamic}. To show this, if the dispatcher pays one dollar per unit service time and $K$ dollars per unit cost (routing and computing), then the price of each computing node is the expected total price that the dispatcher pays to this computing node. In this way, the dispatcher should choose the computing nodes with the lowest prices to ensure the minimization of the cost function. Hence, there exists a threshold for each computing node to decide whether this computing node is activated such that the requests can be scheduled and processed via this computing node. That is, if the price of the computing node is smaller than this threshold, then this computing node will be activated and provide service capacities; otherwise, this computing node will not be activated. For each $i\in\mathcal{N}$ and given any $\theta\geq Kd_{i}$, let $\mathbf{g}_{i}(\theta)$ be the solution to the equation below:
\begin{align}
\label{eqn-8}
K\varphi_{i}(\gamma)+K\gamma\varphi^{\prime}_{i}(\gamma)=\theta.
\end{align}
By detailed computation, $K\varphi_{i}(\gamma)+K\gamma\varphi^{\prime}_{i}(\gamma)$ is larger than $Kd_{i}$ and upper bounded due to $\gamma\leq\gamma^{\max}_{i}$.
Since $\varphi_{i}(\gamma)$ and $\varphi^{\prime}_{i}(\gamma)$ are positive and increasing, $\mathbf{g}_{i}(\cdot)$ is an increasing function. In particular, if $\theta$ exceeds certain bound, which is related to $\gamma=\gamma^{\max}_{i}$ and denoted by $\theta^{\max}_{i}$, then there does not exist any solution to the equation \eqref{eqn-8}. In this case, we define $\mathbf{g}_{i}(\theta)=\mathbf{g}_{i}(\theta^{\max}_{i})$. The following lemma shows the relation between the variable \eqref{eqn-7} and the equation \eqref{eqn-8}.

\begin{lemma}
\label{lem-1}
For each $i\in\mathcal{N}$, there exists $\bar{\gamma}_{i}\in(0, \gamma^{\max}_{i}]$ such that $\theta_{i}=\bar{\gamma}^{-1}_{i}+K\varphi_{i}(\bar{\gamma}_{i})$ and $\bar{\gamma}_{i}=\mathbf{g}_{i}(\theta_{i})$.
\end{lemma}

\begin{IEEEproof}
First, since $\phi$ is convex and $(0, \gamma^{\max}_{i}]$ is bounded, we conclude the existence of $\bar{\gamma}_{i}\in(0, \gamma^{\max}_{i}]$ such that $\theta_{i}=\bar{\gamma}^{-1}_{i}+K\varphi_{i}(\bar{\gamma}_{i})$ and $\bar{\gamma}_{i}$ minimizes the right-hand side of \eqref{eqn-7}. Second, from the derivative of \eqref{eqn-7} and the detailed computation, $\bar{\gamma}^{-1}_{i}=K\bar{\gamma}_{i}\varphi^{\prime}_{i}(\bar{\gamma}_{i})$. From \eqref{eqn-8}, $K\varphi_{i}(\bar{\gamma}_{i})+K\bar{\gamma}_{i}\varphi^{\prime}_{i}(\bar{\gamma}_{i})=K\varphi_{i}(\bar{\gamma}_{i})+\bar{\gamma}^{-1}_{i}=\theta_{i}$, which implies that $\bar{\gamma}_{i}$ is the solution to \eqref{eqn-8} with $\theta=\theta_{i}$. Hence, we conclude that $\bar{\gamma}_{i}=\mathbf{g}_{i}(\theta_{i})$.
\end{IEEEproof}

\begin{theorem}
\label{thm-1}
Consider the optimization problem \eqref{eqn-6}. Let Assumption \ref{asp-1} hold and $\min_{1\leq i\leq n}Kc_{i}<\theta_{1}\leq\ldots\leq\theta_{n}\leq\theta_{n+1}=\max_{1\leq i\leq n}\mathbf{g}^{-1}_{i}(\gamma^{\max}_{i})$ with the inverse function $\mathbf{g}^{-1}_{i}$. The optimal scheduling rates and service capacities are
\begin{align}
\label{eqn-9}
\gamma^{\ast}_{i}&=\left\{\begin{aligned}
&\mathbf{g}_{i}(\theta), && \text{ for } 1\leq i\leq n^{\ast}, \\
&0, && \text{ for } n^{\ast}<i\leq n,
\end{aligned}\right. \\
\label{eqn-10}
u^{\ast}_{i}&=\left\{\begin{aligned}
&\gamma_{i}^{\ast}-\frac{1}{\sqrt{K\varphi^{\prime}_{i}(\gamma^{\ast}_{i})}}, && \text { for } 1\leq i\leq n^{\ast},  \\
&0, && \text { for } n^{\ast}<i\leq n,
\end{aligned}\right.
\end{align}
where $\theta\in(\theta_{n^{\ast}}, \theta_{n^{\ast}+1}]$ is such that $\sum_{i=1}^{n}u_{i}^{\ast}=\lambda$ and
\begin{align}
\label{eqn-11}
n^{\ast}&:=\arg\max_{k\in\mathcal{N}}\left\{\theta_{k}: \sum_{i=1}^{k}\left(\mathbf{g}_{i}(\theta_{k})\right.\right. \nonumber  \\
& \qquad  \left.\left. -\frac{1}{\sqrt{K\varphi^{\prime}_{i}(\mathbf{g}_{i}(\theta_{k}))}}\right)<\lambda\right\}.
\end{align}
\end{theorem}

\begin{IEEEproof}
For the optimization problem \eqref{eqn-6}, all constraints are linear and convex, whereas the cost function is not convex with respect to any pair $(\gamma_{i}, u_{i})$. The Lagrangian for \eqref{eqn-6} is defined as follows:
\begin{align}
\label{eqn-12}
\mathcal{L}&:=\sum_{i=1}^{n}\left(\frac{u_{i}}{\lambda(\gamma_{i}-u_{i})}+\frac{Ku_{i}\varphi_{i}(\gamma_{i})}{\lambda}\right)-\theta\left(\sum_{i=1}^{n}\frac{u_{i}}{\lambda}-1\right) \nonumber \\
&\  -\sum_{i=1}^{n}\mathbf{b}_{i}u_{i}-\sum_{i=1}^{n}\mathbf{h}_{i}(\gamma_{i}-u_{i})+\sum_{i=1}^{n}\mathbf{k}_{i}(\gamma_{i}-\gamma^{\max}_{i}),
\end{align}
where $\theta, \mathbf{b}_{i}, \mathbf{h}_{i}, \mathbf{k}_{i}\in\mathbb{R}^{+}$ are Lagrange multipliers.

Since not all computing nodes need to be activated, we denote by $\mathcal{A}\subset\mathcal{N}$ the set of all inactivated computing nodes, and $u_{i}=\gamma_{i}=0$ for all $i\in\mathcal{A}$. In this respect, the optimization problem \eqref{eqn-6} is rewritten equally as
\begin{subequations}
\label{eqn-13}
\begin{align}
\label{eqn-13a}
\min&\quad \sum_{i\in\mathcal{N}\setminus\mathcal{A}}\lambda^{-1}u_{i}\left((\gamma_{i}-u_{i})^{-1}+K\varphi_{i}(\gamma_{i})\right) \\
\label{eqn-13b}
\text{s.t.}& \quad g_{1}(u, \gamma):=\sum_{i\in\mathcal{N}\setminus\mathcal{A}}u_{i}-\lambda=0,  \\
\label{eqn-13c}
&\quad g_{2i}(u, \gamma):=\gamma_{i}-\gamma^{\max}_{i}\leq0,  \quad \forall i\in\mathcal{N}\setminus\mathcal{A}, \\
\label{eqn-13d}
&\quad g_{3i}(u, \gamma):=u_{i}-\gamma_{i}<0,\quad \forall i\in\mathcal{N}\setminus\mathcal{A},  \\
\label{eqn-13e}
&\quad g_{4i}(u, \gamma):=-u_{i}<0,\quad \forall i\in\mathcal{N}\setminus\mathcal{A}.
\end{align}
\end{subequations}
Next, we need to establish the set $\mathcal{N}\setminus\mathcal{A}$ of all activated computing nodes with optimal scheduling and service rates. From \eqref{eqn-13}, $\nabla g_{1}(u, \gamma)$ is only related to $u$ and linearly independent, whereas $\nabla g_{2}(u, \gamma):=(\nabla g_{21}(u, \gamma), \ldots, g_{2n}(u, \gamma))$ is only related to $\gamma$ and linearly independent. Hence, we can easily check that $\nabla g_{1}(u, \gamma)$ and $\nabla g_{2i}(u, \gamma)$ with $i\in\mathcal{N}\setminus\mathcal{A}$ are linearly independent. From \cite[Definition 4.1]{Hoheisel2009abadie} and \cite[Theorem 4.3]{Hoheisel2009abadie}, the Guignard constraint qualification (GCQ) holds at $(u, \gamma)$, and further from \cite[Theorem 6.1.4]{Bazaraa1976foundations},
the optimal solution (if exists) satisfies the following KKT conditions:
\begin{subequations}
\label{eqn-14}
\begin{align}
\label{eqn-14a}
&\frac{\partial\mathcal{L}}{\partial u_{i}}=\frac{\gamma_{i}}{\lambda(\gamma_{i}-u_{i})^{2}}+\frac{K\varphi_{i}(\gamma_{i})}{\lambda}-\frac{\theta}{\lambda}-\mathbf{b}_{i}+\mathbf{h}_{i}=0, \\
\label{eqn-14b}
&\frac{\partial\mathcal{L}}{\partial\gamma_{i}}=\frac{-u_{i}}{\lambda(\gamma_{i}-u_{i})^{2}}+\frac{Ku_{i}\varphi^{\prime}_{i}(\gamma_{i})}{\lambda}-\mathbf{h}_{i}+\mathbf{k}_{i}=0, \\
\label{eqn-14c}
&\sum_{i=1}^{n}\frac{u_{i}}{\lambda}-1=0, \quad \sum_{i=1}^{n}\mathbf{b}_{i}u_{i}=0,  \\
\label{eqn-14d}
&\sum_{i=1}^{n}\mathbf{h}_{i}(\gamma_{i}-u_{i})=0, \quad \sum_{i=1}^{n}\mathbf{k}_{i}(\gamma_{i}-\gamma^{\max}_{i})=0.
\end{align}
\end{subequations}
It is easy to find that $\mathbf{h}_{i}\equiv0$ holds from $\gamma_{i}>u_{i}$. From \eqref{eqn-14a},
\begin{align}
\label{eqn-15}
u^{\ast}_{i}=\max\left\{0, \gamma^{\ast}_{i}-\sqrt{\frac{\gamma^{\ast}_{i}}{\theta+\lambda\mathbf{b}_{i}-K\varphi_{i}(\gamma^{\ast}_{i})}}\right\},
\end{align}
where $\gamma^{\ast}_{i}$ is the optimal service rate. If $\gamma_{i}<\gamma^{\max}_{i}$, then $\mathbf{k}_{i}=0$ from \eqref{eqn-14d}, and further from \eqref{eqn-14b},
\begin{align}
\label{eqn-16}
u^{\ast}_{i}\left((\gamma^{\ast}_{i}-u^{\ast}_{i})^{-2}-K\varphi^{\prime}_{i}(\gamma^{\ast}_{i})\right)=0.
\end{align}
If $u^{\ast}_{i}>0$, we have from \eqref{eqn-14c} that $\mathbf{b}_{i}=0$, and from \eqref{eqn-15}-\eqref{eqn-16},
\begin{align}
\label{eqn-17}
\gamma^{\ast}_{i}(\gamma^{\ast}_{i}-u^{\ast}_{i})^{-2}&=\theta-K\varphi_{i}(\gamma^{\ast}_{i})=K\gamma^{\ast}_{i}\varphi^{\prime}_{i}(\gamma^{\ast}_{i}),
\end{align}
which shows that $\gamma^{\ast}_{i}=\mathbf{g}_{i}(\theta)$. If $u^{\ast}_{i}=0$ for some $i\in\mathcal{N}$, then the optimal value of $\gamma^{\ast}_{i}$ has no effects on the cost function in \eqref{eqn-6a}, and can be chosen arbitrarily from $[0, \gamma^{\max}_{i}]$.
In this case, we can still choose $\gamma^{\ast}_{i}=\mathbf{g}_{i}(\theta)$, whereas $\mathbf{b}_{i}$ needs to be chosen appropriately to ensure the KKT conditions \eqref{eqn-14a}-\eqref{eqn-14d}. If $\gamma_{i}=\gamma^{\max}_{i}$, then $\mathbf{k}_{i}\geq0$ can be chosen arbitrarily. In this case, we can follow the above argument to derive the same optimal values.

As stated above, if $u^{\ast}_{i}>0$, then $\mathbf{b}_{i}=0$ and from \eqref{eqn-15},
\begin{align}
\label{eqn-18}
\theta>1/\gamma^{\ast}_{i}+K\varphi_{i}(\gamma^{\ast}_{i})\geq\theta_{i},
\end{align}
where $\theta_{i}$ is defined in \eqref{eqn-7}. If $u^{\ast}_{i}=0$, then we have from \eqref{eqn-15} and \eqref{eqn-17} that
\begin{align}
\label{eqn-19}
\theta+\lambda\mathbf{b}_{i}&\leq\mathbf{g}^{-1}_{i}(\theta)+K\varphi_{i}(\mathbf{g}_{i}(\theta)) \nonumber  \\
& \leq\theta-\mathbf{g}_{i}(\theta)(K\varphi^{\prime}_{i}(\mathbf{g}_{i}(\theta))-\mathbf{g}^{-2}_{i}(\theta)).
\end{align}
From $\mathbf{b}_{i}\geq0$, \eqref{eqn-19} holds if either $\mathbf{g}_{i}(\theta)=0$ or $K\varphi^{\prime}_{i}(\mathbf{g}_{i}(\theta))-\mathbf{g}^{-2}_{i}(\theta)\leq0$. In particular, $K\varphi^{\prime}_{i}(\mathbf{g}_{i}(\theta))-\mathbf{g}^{-2}_{i}(\theta)\leq0$ implies that $\mathbf{g}_{i}(\theta)\leq\bar{\gamma}_{i}$, where $\bar{\gamma}_{i}=\mathbf{g}_{i}(\theta_{i})$ is from Lemma \ref{lem-1}. Since $\mathbf{g}_{i}(\cdot)$ is increasing, $\mathbf{g}_{i}(\theta)\leq\mathbf{g}_{i}(\theta_{i})$ implies $\theta\leq\theta_{i}$. Hence, in the derived optimal strategy, the computing nodes are chosen via the increasing order of $\theta_{i}$. In addition, $\sum_{i=1}^{n}u^{\ast}_{i}=\lambda$ needs to be satisfied, thereby resulting in \eqref{eqn-11}.

Finally, we show the uniqueness of the derived optimal solution via the contradiction. Let $(\bar{u}^{\ast}, \bar{\gamma}^{\ast})$ and $(\hat{u}^{\ast}, \hat{\gamma}^{\ast})$ be two different optimal solutions to \eqref{eqn-6}. Since $\sum_{i=1}^{n}\bar{u}^{\ast}_{i}=\sum_{i=1}^{n}\hat{u}^{\ast}_{i}=\lambda$, we have $\bar{u}^{\ast}_{i}\neq\hat{u}^{\ast}_{i}$ for all $i\in\{1, \ldots, \max\{\bar{n}^{\ast}, \hat{n}^{\ast}\}\}$, which further implies from \eqref{eqn-10}-\eqref{eqn-11} that $\bar{\gamma}^{\ast}_{i}\neq\hat{\gamma}^{\ast}_{i}$ for all $i\in\{1, \ldots, n^{\ast}\}$. However, For the following equation
\begin{align}
\label{eqn-20}
\sum_{i=1}^{n}\left(x_{i}-\frac{1}{\sqrt{K\varphi^{\prime}_{i}(x_{i})}}\right)=\lambda,  \quad \forall \lambda>0,
\end{align}
we have from \cite{Fujisawa1971some} that the solution to \eqref{eqn-20} is unique, which implies that $\bar{\gamma}^{\ast}_{i}=\hat{\gamma}^{\ast}_{i}$ for all $i\in\{1, \ldots, n\}$ and results in a contradiction. Therefore, we conclude that the optimal solution to the problem \eqref{eqn-6} (if it exists) is unique.
\end{IEEEproof}

From Theorem \ref{thm-1}, the thresholds are determined first to rearrange all computing nodes and then to decide which computing nodes to be activated. After determining the activated computing nodes, the optimal scheduling and allocation can be established. How to achieve this optimal strategy is summarized in Algorithm \ref{alg-1}. On the other hand, we emphasize that the derived optimal strategy is available for the case that the request arrival is modeled into a random process like Poisson process, and in this case the arrival rate is the rate parameter of Poisson process.

\begin{algorithm}[!t]
\DontPrintSemicolon
\caption{\small{Optimal Scheduling and Allocation}}
\label{alg-1}
\KwIn{$\theta_{i}, i\in\mathcal{N}$}
\KwOut{the optimal scheduling and service rate}
Sort $\theta_{i}, i\in\mathcal{N}$ in a non-decreasing order\;
From \eqref{eqn-11} choose a set of computing nodes to be activated \;
Compute the threshold $\theta\in(\theta_{n^{\ast}}, \theta_{n^{\ast}+1}]$ \;
Set the optimal scheduling using \eqref{eqn-10} \;
Set the optimal allocation using \eqref{eqn-9}
\end{algorithm}

\section{AIMD-based Scheduling}
\label{sec-AIMDschedule}

Due to the constraint \eqref{eqn-6d} in the optimization problem \eqref{eqn-6}, the backlog of each computing node does not need to be considered. However, the backlog of the dispatcher needs to be investigated. In this section, we first apply the AIMD-based strategy to reconsider the request scheduling problem in Section \ref{subsec-AIMDstrategy}, and then propose an approach to determine the AIMD parameters such that the derived optimal strategy is still valid in Section \ref{subsec-paradesign}.

\subsection{AIMD-based Strategy}
\label{subsec-AIMDstrategy}

In the request scheduling, the over-scheduling phenomena may occur. Therefore, $\sum_{i=1}^{n}u_{i}=\lambda$ is imposed in \eqref{eqn-6} to avoid the over-scheduling of the limited requests.
On the other hand, if there exists no backlog in the dispatcher (i.e., $\delta=0$), then it means no enough requests to be scheduled. These two cases may have effects on the derived optimal strategy. In this respect, the scheduling problem needs to be reconsidered. In particular, the scheduling rate may experience a jump in these two cases, which may further affect the service rate. An appropriate method is to model these two cases into an AIMD-based event-triggered system.

In the AIMD strategy, there exist two phases: the additive increase (AI) phase and the multiplicative decrease (MD) phase. In the AI phase, if the scheduling rate of each computing node does no reach its maximum (which is upper bounded via the service rate), then it will increases linearly with an additive rate $\alpha_{i}>0$. Once the maximum is reached, it saturates at this value until there exists no backlog or $\sum_{i=1}^{n}u_{i}=\lambda$. If there exists no backlog in the dispatcher or $\sum_{i=1}^{n}u_{i}=\lambda$, then the scheduling rate needs to be adjusted and experiences an instantaneous decrease with a multiplicative factor $\beta_{i}\in(0, 1)$, which occurs in the MD phase. In these two phases, the MD phase is only activated at certain time instant and results in a jump, after which the AI phase is activated.

In the AIMD mechanism, the occurrence of the MD phase is determined via an event-triggered mechanism (ETM), which depends on the variable $\delta(t)\in\mathbb{R}^{+}$ and the condition $\sum_{i=1}^{n}u_{i}=\lambda$. Let $t_{k}$ be the event-triggering time with $k\in\mathcal{N}$, and $\delta(t_{0})=0$ for the initial time $t_{0}$. Hence, for each computing node, the behavior of its scheduling rate can be mathematically modeled below:
\begin{align}
\label{eqn-21}
u_{i}(t)=\beta_{i}u_{i}(t_{k})+\alpha_{i}(t-t_{k}), \quad \forall t\in(t_{k}, t_{k+1}],
\end{align}
where $i\in\mathcal{N}$ and
\begin{align}
\label{eqn-22}
t_{k+1}:=\min\left\{t>t_{k}: \delta(t)=0 \text{ or } \sum^{n}_{i=1}u_{i}(t)=\lambda\right\}.
\end{align}

From the above discussion, we define $T_{k}:=t_{k+1}-t_{k}$ with $k\in\mathbb{N}$. From \eqref{eqn-1} and \eqref{eqn-21}-\eqref{eqn-22}, we have that $\delta=0$ equals to
\begin{align}
\label{eqn-23}
\lambda T_{k}=\int_{t_{k}}^{t_{k+1}}u(t)dt=\sum_{i=1}^{n}(\beta_{i} u_{i}(t_{k})+0.5\alpha_{i}T_{k})T_{k},
\end{align}
which is similar to the one in \cite{Vlahakis2021AIMD}. Comparing \eqref{eqn-23} and the condition $\sum_{i=1}^{n}u_{i}=\lambda$ with \eqref{eqn-21}, we find that \eqref{eqn-23} will not be activated for the case where $\lambda$ is fixed. In this case, the ETM \eqref{eqn-22} can be written equivalently as
\begin{align}
\label{eqn-24}
t_{k+1}&=\min\left\{t>t_{k}: \lambda=\sum_{i=1}^{n}(\beta_{i}u_{i}(t_{k})+\alpha_{i}(t-t_{k}))\right\},
\end{align}
which implies that the backlog would not be zero. However, if $\lambda$ switches among different modes or additional constraints are imposed on the backlog $\delta$, then the condition \eqref{eqn-23} or its variants should be included into the ETM \eqref{eqn-22}, which deserves further study. For the request scheduling with the AIMD-based strategy, we have the following proposition to ensure the convergence of scheduling rates.

\begin{proposition}
\label{prop-1}
If the request scheduling follows the AIMD-based strategy, then for all $i\in\mathcal{N}$, the scheduling rate of each computing node converges to a point given below:
\begin{align}
\label{eqn-25}
\mathbf{u}^{\ast}_{i}:=\min\{(1-\beta_{i})^{-1}\alpha_{i}T^{\ast}, \gamma_{i}-\epsilon\},
\end{align}
where $\epsilon>0$ is sufficiently small, $\gamma_{i}\in\mathbb{R}^{+}$ is the service rate, and the constant $T^{\ast}>0$ is such that
\begin{align}
\label{eqn-26}
\sum_{i=1}^{n}\mathbf{u}^{\ast}_{i}=\lambda.
\end{align}
\end{proposition}

The proof of Proposition \ref{prop-1} follows the similar mechanism as in Lemma IV.2 of \cite{Studli2015modified}, and is omitted here. The differences lies in that the lower bound of the scheduling rate is not considered here and the upper bound is not a constant given \emph{a priori} but the service rate which is derived from Theorem \ref{thm-1}. Therefore, $\mathbf{u}^{\ast}_{i}$ in \eqref{eqn-25} is not fixed but related to the service rate, and the introduction of $\epsilon>0$ comes from $\gamma_{i}>u_{i}$ in \eqref{eqn-6d}. In this respect, $T^{\ast}$ can be computed via an iterative method summarized in Algorithm \ref{alg-2}. To be specific, given a initial value of $T^{\ast}$ as in line 1 of Algorithm \ref{alg-2}, we compute the scheduling rate $\mathbf{u}^{\ast}_{i}$ as in line 2. If $\lambda-\sum_{i=1}^{N}\mathbf{u}^{\ast}_{i}>\varepsilon$ with a sufficiently small threshold $\varepsilon>0$, then both $T^{\ast}$ and $\mathbf{u}^{\ast}_{i}$ are computed iteratively as in lines 3-4. This iteration is terminated until $\lambda-\sum_{i=1}^{N}\mathbf{u}^{\ast}_{i}$ reaches the threshold $\varepsilon>0$. On the other hand, if $\mathbf{u}^{\ast}_{i}=\alpha_{i}(1-\beta_{i})^{-1}T^{\ast}$ in \eqref{eqn-25}, then
\begin{align}
\label{eqn-27}
T^{\ast}=\lambda\left(\sum^{n}_{i=1}(1-\beta_{i})^{-1}\alpha_{i}\right)^{-1}.
\end{align}

\begin{algorithm}[!t]
\DontPrintSemicolon
\caption{Computation of $T^{\ast}$}
\label{alg-2}
\KwIn{the constants $\lambda, \alpha_{i}, \beta_{i}$ for $i\in\mathcal{N}$ and the threshold $\varepsilon>0$}
\KwOut{$\mathbf{u}^{\ast}=(\mathbf{u}^{\ast}_{1}, \ldots, \mathbf{u}^{\ast}_{n})$ and $T^{\ast}$}
Set $T^{\ast}_{0}=\lambda\Delta^{-1}$ with $\Delta:=\sum^{n}_{i=1}\frac{\alpha_{i}}{1-\beta_{i}}$ \;
For all $i\in\mathcal{N}$, compute $\mathbf{u}^{\ast}_{i}$ as in \eqref{eqn-25} \;
\While{$\lambda-\sum_{i=1}^{n}\mathbf{u}^{\ast}_{i}>\varepsilon$}{
Compute $T^{\ast}=T^{\ast}_{0}+(\lambda-\sum^{n}_{i=1}\mathbf{u}^{\ast}_{i})\Delta^{-1}$ \;
For all $i\in\mathcal{N}$, compute $\mathbf{u}^{\ast}_{i}$ as in \eqref{eqn-25} \;
$T^{\ast}_{0}=T^{\ast}$ \;
}
\end{algorithm}

\subsection{AIMD Parameters Design}
\label{subsec-paradesign}

For each scheduling rate obtained from the AIMD-based strategy, the service rate determines the convergent point of each scheduling rate. From Theorem \ref{thm-1}, both the scheduling rate and the service rate are optimized. A direct way is to combine the solutions in Theorem \ref{thm-1} and Proposition \ref{prop-1} such that the convergence of all scheduling rates is to the optimal scheduling rates, thereby balancing the resource allocation mechanism and the AIMD-based strategy.

From Theorem \ref{thm-1}, not all computing nodes are activated. For these inactivated computing nodes, both $u_{i}$ and $\gamma_{i}$ are zero. Hence, we can set $\alpha_{i}=\beta_{i}=0$ simply. In the following, we only investigate all activated computing nodes, whose scheduling rates and service capacities are given below with $i\in\{1, \ldots, n^{\ast}\}$:
\begin{align}
\label{eqn-28}
\gamma^{\ast}_{i}&=\mathbf{g}_{i}(\theta),  \quad u^{\ast}_{i}=\gamma_{i}^{\ast}-\frac{1}{\sqrt{K\varphi^{\prime}_{i}(\gamma_{i}^{\ast})}}.
\end{align}
In terms of scheduling rates, $u^{\ast}_{i}$ and $\mathbf{u}^{\ast}_{i}$ should be equivalent. In this case, from \eqref{eqn-25} and \eqref{eqn-28}, we have
\begin{align}
\label{eqn-29}
\gamma_{i}^{\ast}-\frac{1}{\sqrt{K\varphi^{\prime}_{i}(\gamma_{i}^{\ast})}}=\min\left\{\frac{\alpha_{i}T^{\ast}}{1-\beta_{i}}, \gamma^{\ast}_{i}-\epsilon\right\}.
\end{align}
If $(1-\beta_{i})^{-1}\alpha_{i}T^{\ast}>\gamma^{\ast}_{i}-\epsilon>0$, then from \eqref{eqn-29},
\begin{align*}
\frac{1}{\sqrt{K\varphi^{\prime}_{i}(\gamma_{i}^{\ast})}}=\epsilon.
\end{align*}
Since $\epsilon>0$ is arbitrarily small, we have from \eqref{eqn-8} and the derivative of $\varphi_{i}$ that $\gamma^{\ast}_{i}$ needs to be sufficiently large, which contradicts with the boundedness of $\gamma_{i}$. Therefore, $(1-\beta_{i})^{-1}\alpha_{i}T^{\ast}\leq\gamma^{\ast}_{i}-\epsilon$, and from \eqref{eqn-26},
\begin{align}
\label{eqn-30}
\gamma_{i}^{\ast}-\frac{1}{\sqrt{K\varphi^{\prime}_{i}(\gamma_{i}^{\ast})}}&=\frac{\alpha_{i}T^{\ast}}{1-\beta_{i}},
\end{align}
which is required to be satisfied by the parameters $\alpha_{i}$ and $\beta_{i}$ for all activated computing nodes. From \eqref{eqn-30}, the choices of $\alpha_{i}, \beta_{i}$ for all activated computing nodes affect each other. In particular, if $\alpha_{i}T^{\ast}$ is treated as a single variable, then we can derive the relation between $\beta_{i}$ and $\alpha_{i}T^{\ast}$. Once $\beta_{i}$ and $\alpha_{i}T^{\ast}$ are determined, we can further derive $\alpha_{i}$ and $T^{\ast}$ explicitly.

\subsection{Further Discussion}
\label{subsec-discuss}

From Section \ref{subsec-AIMDstrategy}, the backlog in the dispatcher is discussed, while is not involved in the ETM \eqref{eqn-24}. Therefore, with the applied AIMD-based mechanism, the backlog in the dispatcher would be unbounded. In this respect, we discuss the backlog in the dispatcher further. From \cite[Section 4.3]{Presti1996bounds} where the arrival rate is modeled to be time-varying, the upper bound of the backlog in the dispatcher can be approximated by some function, and the convergence of this function implies the convergence of the backlog in the dispatcher. For the fixed arrival rate in this paper, this function is not convergent, which further results in the reconsideration of the arrival rate.

In the following, we propose an alternative method to adjust the arrival rate while having not effects on the request scheduling and resource allocation afterwards. Let $\delta_{\min}\geq0$ and $\delta_{\max}>\delta_{\min}$ be two thresholds for the variable $\delta(t)$. If $\delta(t)=\delta_{\min}$, then the backlog in the dispatcher reaches its minimum such that the following scheduling and allocation are affected. If $\delta(t)=\delta_{\max}$, then the backlog in the dispatcher reaches its maximum such that the arrival rate should be decreased. In these two cases, the dispatcher sends a warning signal to the request arrival such that the arrival rate switches in two modes, which is summarized below.
\begin{align}
\label{eqn-31}
\bar{\lambda}:=\left\{\begin{aligned}
&\lambda, &\quad & \delta(t)\leq\delta_{\min}, \\
& \rho\lambda, &\quad & \delta(t)\geq\delta_{\max},
\end{aligned}\right.
\end{align}
where $\rho\in(0, \min_{i\in\mathcal{N}}\{\beta_{i}\})$ is constant, $\bar{\lambda}$ is the real arrival rate, and $\lambda$ is the desired arrival rate for the scheduling and allocation. Here, the constraint on $\rho$ is to ensure the decrease of the backlog in the dispatcher; see also \eqref{eqn-1}. From \eqref{eqn-31}, if $\delta(t)=\delta_{\min}$, then the arrival rate is in the first mode until $\delta(t)$ increases to its maximum $\delta_{\max}$ due to the limited capacity of the dispatcher; if $\delta(t)=\delta_{\max}$, then the arrival rate is in the second mode until $\delta(t)$ decreases to its minimum $\delta_{\min}$ due to the optimal strategy in Section \ref{sec-optimalschedul} and \eqref{eqn-26}. Therefore, the backlog in the dispatcher is bounded in $[\delta_{\min}, \delta_{\max}]$. By choosing $\delta_{\min}$ and $\delta_{\max}$ appropriately, the following scheduling and allocation strategies in Sections \ref{sec-optimalschedul} and \ref{sec-AIMDschedule} will not be affected. In particular, in the second mode, due to the backlog in the dispatcher, the decrease of the arrival rate has no effects on the scheduling and allocation strategies. Hence, such switching mechanism guarantees both the derived strategy and the boundedness of the backlog in the dispatcher. A general case is that the arrival rate switches in different modes having effects on the scheduling and allocation strategies, which deserves further studies.

\begin{figure}[!t]
\begin{center}
\begin{picture}(65, 95)
\put(-65, -15){\resizebox{65mm}{40mm}{\includegraphics[width=2.5in]{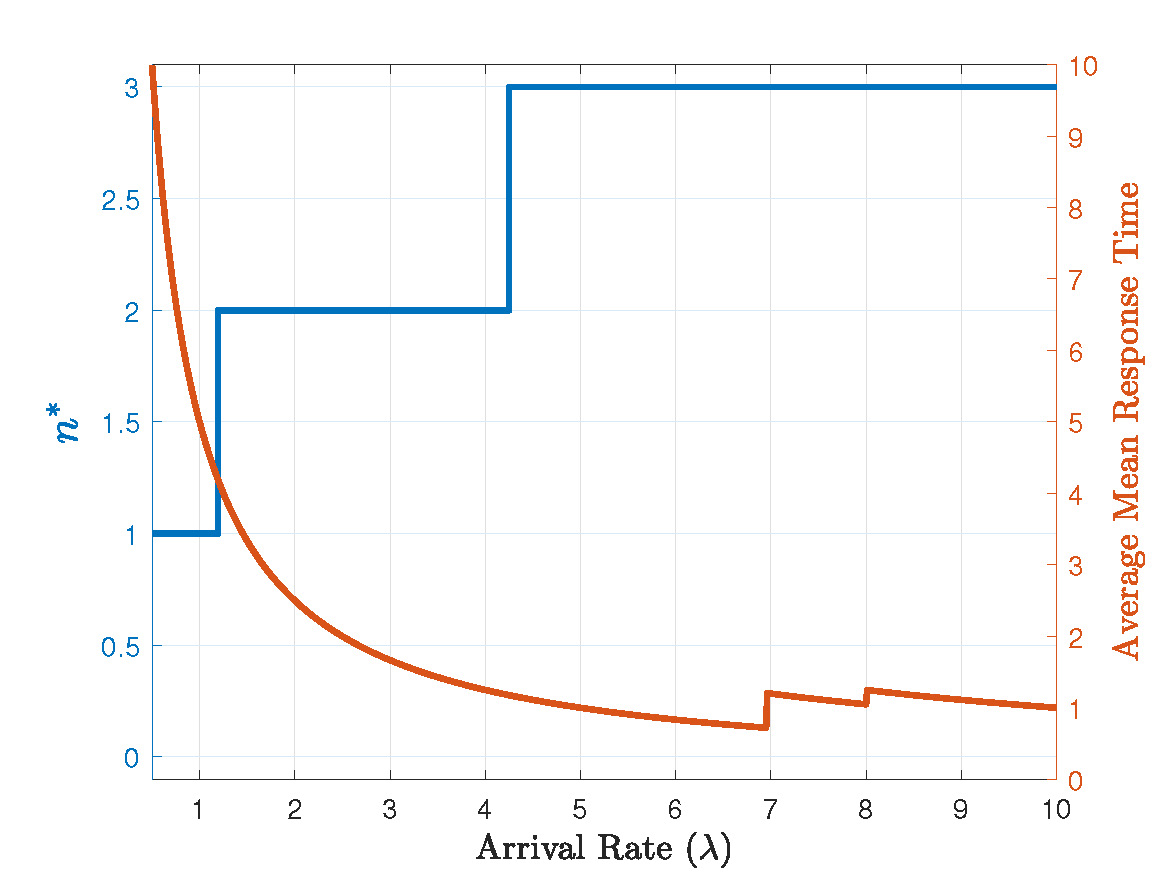}}}
\end{picture}
\end{center}
\caption{Evolution of the optimal number $n^{\ast}$ in \eqref{eqn-11} and the average mean response time $\mathcal{T}$ in \eqref{eqn-2} with the arrival rate $\lambda$.}
\label{fig-2}
\end{figure}

\section{Numerical Example}
\label{sec-example}

In this section, a numerical example is presented to illustrate the derived results. We consider a distributed computing system with 3 computing nodes with different service capacities. These heterogeneous computing nodes have different power consumption and processing capacities. Typically, three types are lightweight computing node (denoted by Node 1) with low service rate and power consumption, middleweight computing node (denoted by Node 2) with medium service rate and power consumption, heavyweight computing node (denoted by Node 3) with high service rate and power consumption. That is, in \eqref{eqn-4}, we set $a_{1}=0.1, a_{2}=0.2, a_{3}=0.5, b_{1}=b_{2}=b_{3}=2, c_{1}=0.2, c_{2}=0.5, c_{3}=0.7, d_{1}=1, d_{2}=2, d_{3}=5$. In addition, $\gamma^{\max}_{1}=5, \gamma^{\max}_{2}=6, \gamma^{\max}_{3}=8$ and the weight $K=1$. Therefore, from \eqref{eqn-7} we can compute that $\theta_{1}=2.1898, \theta_{2}=3.6978, \theta_{3}=7.1296$. Hence, when the requests come, the lightweight computing node is activated first. Due to the boundedness of $\gamma_{i}$ with $i\in\{1, 2, 3\}$, we have the upper bound of $\theta_{i}$, that is, $\theta_{4}=112.2$. Here, we choose $\theta=11.648$ and $\lambda=8$, and from Theorem \ref{thm-1}, we can derive the optimal scheduling rates $u^{\ast}_{1}=4.4393, u^{\ast}_{2}=2.5180, u^{\ast}_{3}=1.0428$ and service rate $\gamma^{\ast}_{1}=5.3281, \gamma^{\ast}_{2}=3.2623, \gamma^{\ast}_{3}=1.6897$. Note that the number of all activated computing nodes depends on the arrival rate. With different arrival rates, the number of the activated computing nodes is shown in Fig. \ref{fig-2}. In addition, the evolution of the average mean response time in \eqref{eqn-2} is also presented in Fig. \ref{fig-2}. From Fig. \ref{fig-2}, with the increase of the arrival rate, the number of the activated computing nodes increases, whereas the average mean response time decreases.

\begin{figure}[!t]
\begin{center}
\begin{picture}(65, 95)
\put(-65, -15){\resizebox{65mm}{40mm}{\includegraphics[width=2.5in]{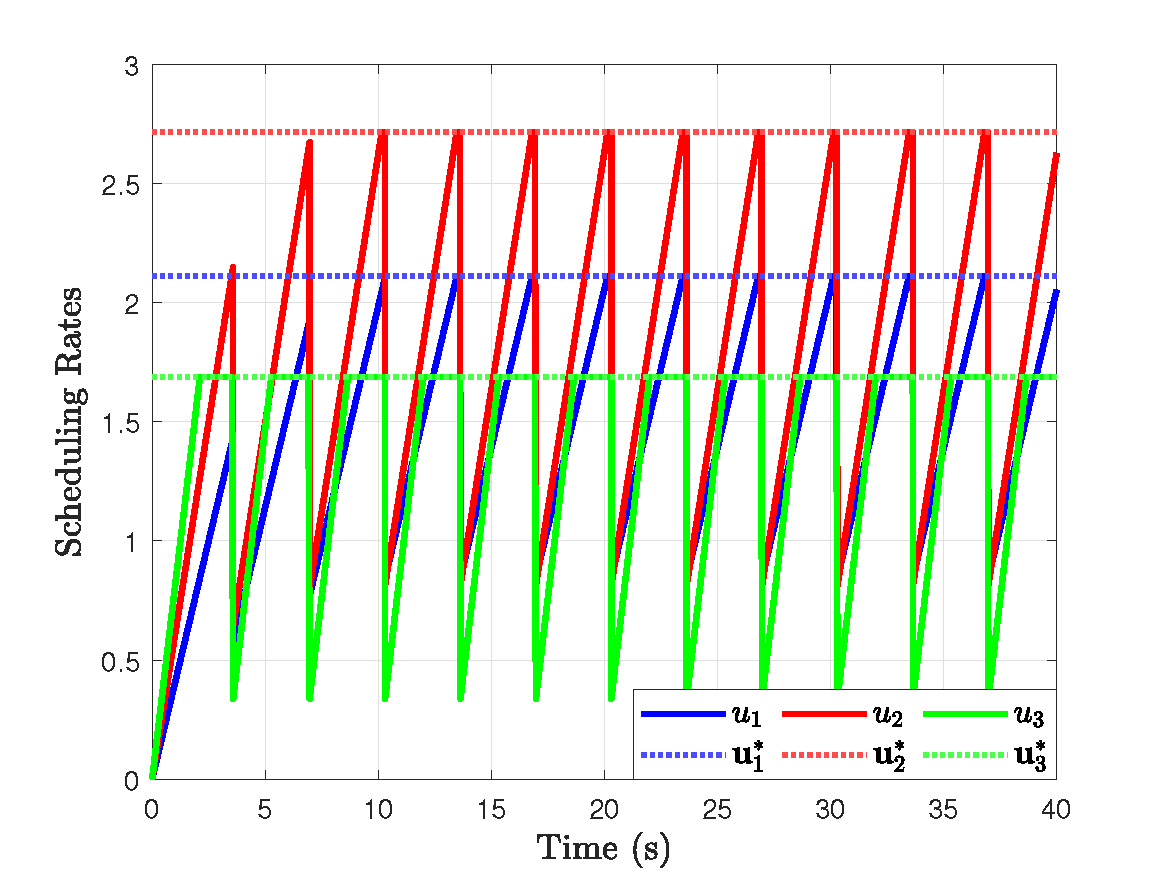}}}
\end{picture}
\end{center}
\caption{Evolution of the scheduling rates $u_{i}$ with the time, $i=\{1, 2, 3\}$. The dashed line is the scheduling rates obtained from \eqref{eqn-25}.}
\label{fig-3}
\end{figure}

Next, we consider the AIMD-based strategy. Let $\lambda=8, \alpha_{1}=0.4, \alpha_{2}=0.6, \alpha_{3}=0.8, \beta_{1}=0.4, \beta_{2}=0.3, \beta_{3}=0.2$ and $\epsilon=0.001$. We can compute from \eqref{eqn-29} that $T^{\ast}=3.1698$, and further from Proposition \ref{prop-1} that $\mathbf{u}^{\ast}_{1}=2.1132, \mathbf{u}^{\ast}_{2}=2.7170, \mathbf{u}^{\ast}_{3}=1.6887$. It is easy to check that $\mathbf{u}^{\ast}_{1}+\mathbf{u}^{\ast}_{2}+\mathbf{u}^{\ast}_{3}\leq\lambda$. In addition, we can see that given the AIMD parameters, the scheduling rates from \eqref{eqn-25} are different from those from Theorem \ref{thm-1}. In particular, $\mathbf{u}^{\ast}_{i}=(1-\beta_{i})^{-1}\alpha_{i}T^{\ast}$ for Nodes 1 and 2, whereas $\mathbf{u}^{\ast}_{3}=\gamma^{\ast}_{3}-\epsilon$ for Node 3. Under the AIMD-based strategy, the evolution of the scheduling rates is presented in Fig. \ref{fig-3}. Note that the scheduling rates of all computing node reach their upper bounds during the AI phase and then do not further increase.

\begin{figure}[!t]
\begin{center}
\begin{picture}(65, 95)
\put(-65, -12){\resizebox{65mm}{40mm}{\includegraphics[width=2.5in]{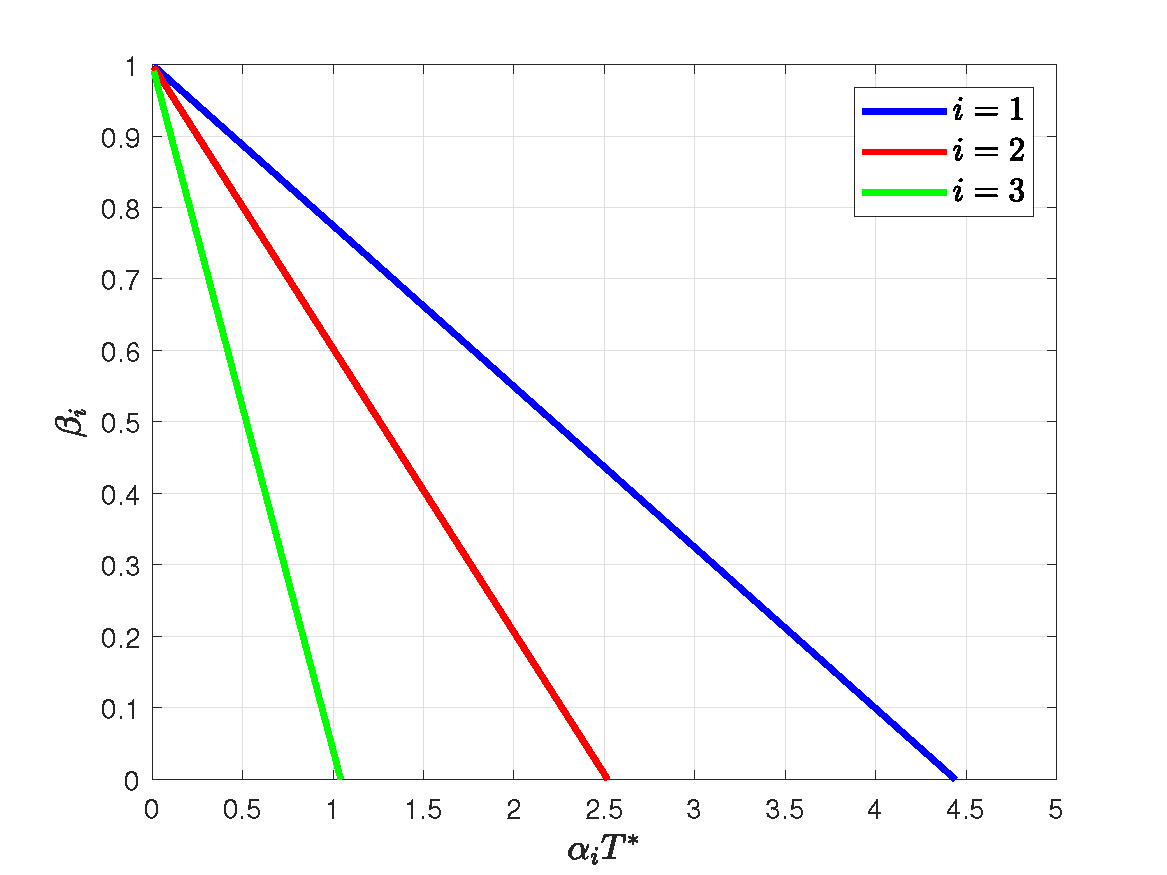}}}
\end{picture}
\end{center}
\caption{The relation $\alpha_{i}T^{\ast}$ and $\beta_{i}$ from \eqref{eqn-29}.}
\label{fig-4}
\end{figure}

Finally, if the AIMD parameters are unknown \emph{a priori}, then we can design these parameters via the relation in \eqref{eqn-29}. In particular, if we consider $\alpha_{i}T^{\ast}$ as a single variable, we can derive the relation between $\beta_{i}$ and $\alpha_{i}T^{\ast}$; see Fig. \ref{fig-4}. That is, if we choose $\beta_{i}\in(0, 1)$ \emph{a priori}, then from Fig. \ref{fig-4} the corresponding value of $\alpha_{i}T^{\ast}$ is determined. The value of $\alpha_{i}$ can be designed such that \eqref{eqn-30} and \eqref{eqn-26} are satisfied.

\section{Conclusion}
\label{sec-conclusion}

In this paper, we investigated the problems of cost-aware request scheduling and resource allocation for distributed computing systems. By proposing a cost function of the response time and service cost, we proposed an optimal scheduling and allocation strategy to minimize the cost function. We further considered the AIMD mechanism to model the relation between the incoming requests and the request scheduling, and combined the AIMD mechanism and the derived optimal strategy to determine the AIMD parameters. Finally, the derived results were illustrated via a numerical example. Future research will be directed to general cases involving in switching arrival rates and different request priorities.


\bibliographystyle{IEEEtran}

\end{document}